\documentclass[aps,prl,twocolumn,amsmath,amssymb,superscriptaddress,
showpacs]{revtex4}

\usepackage{graphicx}
\usepackage{dcolumn}
\usepackage{bm}
\usepackage{ifthen}
\usepackage{booktabs}

\begin{document}

\title{Highly-reduced Fine-structure splitting in InAs/InP quantum dots 
offering efficient on-demand 1.55 $\mu$m entangled photon emitter}
\author{Lixin He \footnote{corresponding author: helx@ustc.edu.cn}}
\affiliation{Key Laboratory of Quantum Information, 
University of Science and Technology
of China, Hefei, 230026, People's Republic of China}
\author{Ming Gong}
\affiliation{Key Laboratory of Quantum Information, 
University of Science and Technology
of China, Hefei, 230026, People's Republic of China}
\author{Chuan-Feng Li}
\affiliation{Key Laboratory of Quantum Information, 
University of Science and Technology
of China, Hefei, 230026, People's Republic of China}
\author{Guang-Can Guo}
\affiliation{Key Laboratory of Quantum Information, 
University of Science and Technology
of China, Hefei, 230026, People's Republic of China}
\author{Alex Zunger}
\affiliation{National Renewable Energy Laboratory, Golden, Colorado,
  80401,USA}

\date{\today }

\begin{abstract}

To generate entangled photon pairs via quantum dots (QDs), 
the exciton fine structure splitting (FSS) must be 
comparable to the exciton homogeneous line width.  
Yet in the (In,Ga)As/GaAs QD, the intrinsic FSS is about a few tens $\mu$eV.
To achieve photon entanglement, it is 
necessary to  Cherry-pick a  sample with extremely small FSS from a large
number of samples, or to apply strong 
in-plane magnetic field. 
Using theoretical modeling of the fundamental causes of FSS
in QDs, we predict that the intrinsic FSS of
InAs/InP QDs is an order of magnitude smaller than that of
InAs/GaAs dots, and better yet, 
their excitonic gap matches the 1.55 $\mu$m fiber optic wavelength,
therefore offer
efficient on-demand entangled photon emitters for long distance 
quantum communication.

\end{abstract}



\maketitle

\begin{figure}
\begin{center}
\includegraphics[width=2.8in]{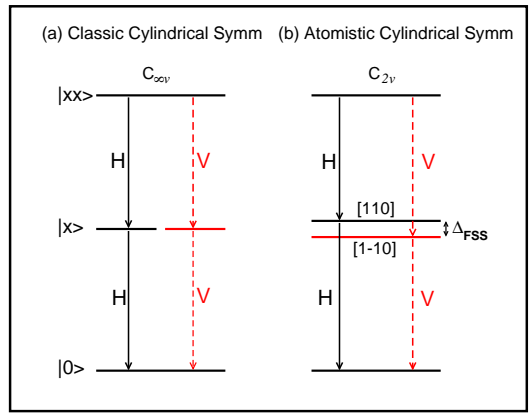}
\end{center}
\caption{ (a) A schematic illustration of the biexciton cascade process to
  generate polarization entangled photons in a QD with classic cylindrical symmetry.
  $H$ and $V$ denote the polarization along the [110] and [1$\bar{1}$0]
  direction, respectively.
(b) Due to the in-plane asymmetry, the $H$ and $V$ polarized
photons have a small energy splitting $\Delta_{\text{FSS}}$, which may destroy
  the polarization entanglement.}
\label{fig:biexciton}
\end{figure}

Entangled photon pairs distinguished themselves
from the classically correlated photons 
because of their non-locality\cite{ou88,aspect82} and therefore
play a crucial role in quantum information
applications, including
quantum teleportation\cite{jennewein01}, quantum
cryptography\cite{gisin02} and 
distributed quantum computation\cite{cirac99}, etc. 
Benson {\it et al.}\cite{benson00} proposed that a biexciton cascade process
in a self-assembled QD can be 
used to generate the ``event-ready'' entangled photon pairs, 
with orders of magnitude higher efficiency than the
traditional parametric down conversion method
\cite{ou88,jennewein01, gisin02}.
This process is shown schematically in
Fig. \ref{fig:biexciton}(a), in which 
a biexciton decays into two photons via two paths of different
polarizations $|H\rangle$ and $|V \rangle$. If the two
paths are indistinguishable, the final result is a
polarization entangled photon pair state\cite{stevenson06, 
benson00}$(|H_{xx} H_{x}\rangle +  |V_{xx} V_{x}\rangle) /\sqrt{2}$.
However, early attempts\cite{benson00} 
to generate the entangled photon pairs using
the InAs/GaAs QDs were unsuccessful, because the $|H\rangle$-
and $|V\rangle$-polarized 
photons have a small energy 
difference due to the asymmetric electron-hole 
exchange interaction in the QDs [see Fig. 1(b)]. 
The small energy splitting, known as the fine structure splitting (FSS),
is typically about -40 $\sim$ +80 $\mu$eV in the InAs/GaAs QDs
\cite{young05,hogele04,tartakovskii04},  which is much larger than the
radiative linewidth ($\sim$ 1.0 $\mu$eV)
\cite{stevenson06, akopian06}. 
Such a splitting provides therefore ``which way'' information about the photon
decay path that can destroy the photon entanglement,  
leaving only classically correlated photon pairs
\cite{stevenson06,akopian06}.

To achieve photon entanglement, 
the FSS must be reduced to a value comparable to 
the exciton homogeneous line width. 
Lack of detailed understanding of the factors controlling FSS in QDs has
thus far impeded the design of small FSS values in quantum systems.
It was, however, empirically discovered that the FSS in InAs/GaAs QDs
can be significantly reduced or even reversed by various thermal annealing protocols 
\cite{young05,tartakovskii04, seguin05}. Furthermore, in alloy dots of 
(In,Ga)As/GaAs, different random realizations of Ga and In distributions on the cation
lattice lead a distribution of FSS values. Carefully screening of dots out of a large
ensemble can then be used to \cite{stevenson06}
found those with small FSS. By applying such ``Cherry Picking''
techniques, 
entangled photon pairs have been recently achieved in the 
(In,Ga)As/GaAs QDs\cite{stevenson06}.
However, even after thermal annealing, 
the FSS is still about $\pm$ 10 $\mu$eV, too large for 
generating entangled photon pairs\cite{stevenson06}. 
The FSS can be further reduced by applying an in-plane magnetic 
field \cite{stevenson06}, which, however, significantly complicates the
experimental setup.
Other methods, e.g. applying in plane electric field, can also 
reduce the FSS\cite{gerardot07}. However, the ensuing Stark effect greatly suppress 
the photoluminescence (PL) intensity\cite{gerardot07}.
Furthermore, after thermal annealing, the 
exciton wavelength is reduced to  880 $\sim$ 950 nm,
becoming uncomfortably close to the energy of the wetting layer emission,
thus leading to unwanted strong background light\cite{young06, stevenson06},
and the wavelength is also too short for the optical fiber communications.

The reason for the early optimism about the use of QD for generating entangled photon
pairs stems from the thought that FSS of an exciton will vanish [see Fig. 
 \ref{fig:biexciton}(a)] in shape-symmetric dots (e.g, cylindrical, or lens-shaped or 
cone-shaped)\cite{bayer99}. However, the FSS of an 
exciton in a dot contains two terms:
the previously largely ignored ``intrinsic  FSS''\cite{bayer99}, 
which is nonzero even in a shaped-symmetric dot, 
and the ``shape asymmetric FSS'' due to deviation from geometric symmetry
along [110] and [1$\bar{1}$0] directions (e.g, lens-shaped with non-circular
base)\cite{seguin05}.
(Note that in this sense, the FSS is just like the
spin-splitting effects, which are composed of the intrinsic Dresselhaus term, due to
the bulk inversion asymmetry and the Rashba term, due to the
geometrical asymmetry.) Whereas the contribution to the FSS of 
QD shape-asymmetric\cite{seguin05} can be reduced by carefully controlling the 
growth conditions, by annealing, or by cherry-picking the ``right dots'' out of 
an ensemble, the ``intrinsic'' FSS is still present
even for an idea cylindrical dot, becasue 
semiconductor materials from which dots are commonly made
have the well-known zinc-blende structure,
and are thus not spatially isotropic. 
The zinc-blende structure has T$_d$ symmetry, so even a cylindrically-shaped, 
i.e., lens or cone QD made of a  zinc-blende semiconductor can only 
have a subgroup C$_{2v}$ symmetry\cite{bester05a}. 
Since the interface between the dot material and the surrounding matrix
material is not necessary a reflection plane, 
the (atomistic) potentials are different along the [110] and [1$\bar{1}$0]
directions\cite{bester05a}. 
This interfacial asymmetry leads to a natural, build-in ``intrinsic FSS''
[Fig. \ref{fig:biexciton}(b)] (even if atoms are unrelaxed, fixed to their ideal 
zinc-blende lattice sites). Such atomistic effects are commonly missed by continuum models
(such as the effective mass approximation and the few-band 
k$\cdot$p method), which only ``see'' the macroscopic shape symmetry, rather than
atomistic details\cite{zunger01}. The intrinsic FSS is thus missed by these 
``low resolution'' methods. However, recent atomistic calculations show 
that the intrinsic FSS is around several tens of
$\mu$eV in the InAs/GaAs QDs \cite{bester03}.

\begin{figure}
\begin{center}
\includegraphics[width=3.0in]{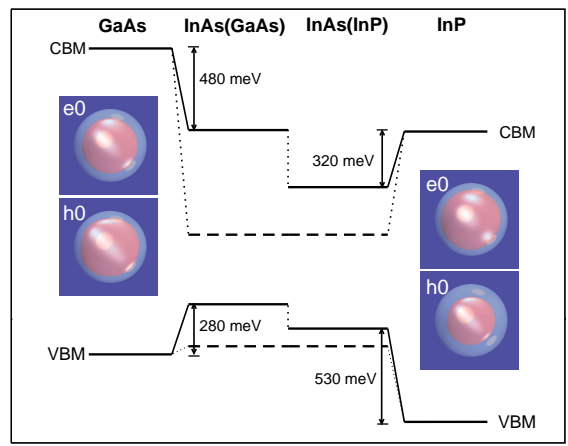}
\end{center}
\caption{
Energy bandoffsets for the InAs/GaAs and InAs/InP QDs. 
The solid lines represent the strain modified bandoffsets,
whereas the dashed line is the unstrained CBM and VBM
for InAs. 
We show in the insert, the first electron (e0) and hole (h0)
wavefunctions of the lens-shaped InAs/GaAs and InAs/InP QDs, 
with base= 25 nm,
and height= 2.5 nm. The isosurface is chosen to enclose 95 \%
of the total density.}
\label{fig:bandoffset}
\end{figure}

Here we design a QD system with reduced FSS by using the 
microscopic understanding of the origins of the intrinsic FSS \cite{bester03}. 
We recognize three factors here. 
First, atomic relaxation due to lattice size-mismatch between
the dot and matrix materials (e.g, InAs and GaAs, respectively, have a 7\% mismatch)
enhances the magnitude of both intrinsic and shape asymmetry FSS\cite{bester03}. Thus one could
expect that the dot/matrix system with lower lattice mismatch should have smaller FSS.
Second, a stronger confinement of the electron and hole in their respective potential
wells will reduce the penetration of the respective wavefunctions into the
matrix material, thus reducing their amplitude at the interface, where intrinsic 
asymmetry is present. Third, since the (atomistic) hole wavefunctions are more
localized on the anion sites, having  dot and matrix material with different anions 
(e.g, As vs P or N) will further reduces the amplitude of hole wavefunctions at
the interface, thus reducing the intrinsic FSS. 

Surprisingly, all three conditions can be met by retaining 
InAs as the dot material, but replacing
the commonly used GaAs matrix material by InP. 
The latter material has a smaller lattice mismatch
with InAs (3 \% instead of 7 \% for GaAs), and manifests a different anion (P)
with respect to the As atom in GaAs. To examine the extent to which the 
InP matrix better confines the excitonic wavefunction 
inside the InAs dot, we compare
in Fig. \ref{fig:bandoffset} the strain-modified potential in an InAs 
dot surrounded by GaAs or InP. 
Since the electron wavefunction (e0 in the insert to Fig.\ref{fig:bandoffset} ) is
similar in both systems due to the light electron mass of InAs, the
dimension of the excitonic wavefunction is mainly determined by the hole wavefunction.
As shown in Fig. \ref{fig:bandoffset}, the (strained) 
hole confining potential for
InAs/InP is 530 meV, much larger than that for InAs/GaAs (46 meV).
Consequently, 
the hole wavefunction of the InAs/InP system (h0 in the insert to
Fig. \ref{fig:bandoffset}) is indeed much more localized  
in the dot interior than is the case in 
the InAs/GaAs dot\cite{gong08a}.
The reduce wavefunction amplitude at the interface,
where the [110] vs [1$\bar{1}$0] asymmetry is maximal,
will then reduce the intrinsic
FSS in InAs/InP.

\begin{table}
\caption{Geometries of the QDs used in the calculations. $D$ is the
  base diameter of the lens-shaped and (truncated-)cone-shaped dots, whereas
  $h$ is the height of the dots. $S$ is defined as $R_{[110]}\cdot
  R_{[1\bar{1}0]}$, where $R_{[110]}$ and $R_{[1\bar{1}0]}$ 
are the diameters of the (elongated) QDs
along [110] and [1$\bar{1}$0] direction, respectively.}
\begin{tabular}{c|c|c} 
\hline \hline
       &  Shape  & Size   \\
\hline
L1   & Lens & $D$ = 20 nm, $h$ = 2.5 to 5.5 nm \\
L2   & Lens & $D$ = 25 nm, $h$ = 2.5 to 5.5 nm \\
L3   & Lens & $h$ = 3 nm, $D$ = 20 to 25  nm \\
L4   & Lens & $h$ = 4 nm, $D$ =  20  to 25  nm \\
L5   & Lens & $h$ = 5 nm, $D$ = 20 to 25  nm \\
C1   & Cone & $D$ = 20 nm, $h$ =  2.5 to 5.5 nm \\
C2   & Cone & $D$ = 25 nm, $h$ = 2.5 to 5.5 nm \\
C3   & Cone & $h$ = 3 nm, $D$ = 20 to 25  nm \\
C4   & Cone & $h$ = 4 nm, $D$ = 20 to 25  nm \\
C5   & Cone & $h$ = 5 nm, $D$ = 20 to 25  nm \\
E1   & Elongated & $S=20^2$ nm$^2$, $h=4.5$ nm \\
E2   & Elongated & $S=25^2$ nm$^2$, $h=4.5$ nm \\ \hline
\end{tabular}
\label{tab:geometry}
\end{table}

To examine our design principles for reduced FSS at work, we carried out 
extensive calculations of the exciton energies and their FSS for 184  
different dots in the InAs/InP and InAs/GaAs QDs. 
We have considered realistic sizes and geometries, including lens-shaped
QDs (L1-L5), (truncated-)cone-shaped QDs (C1-C5)
and elongated QDs (E1,E2) 
(see Table~\ref{tab:geometry}).
We use an
atomistic pseudopotential approach to 
describe the single-particle physics\cite{williamson00,gong08a},
and a configuration-interaction approach to describe the many-body
interactions\cite{franceschetti99}.
The atomic positions were
relaxed to their minimum-strain position. Since the exciton and biexciton are
nearly linearly polarized along the[110] direction and the[1$\bar{1}$0]direction, the
FSS is defined as the energy splitting between the[110] polarized exciton
and[1$\bar{1}$0]polarized exciton, i.e., $\Delta_{\rm FSS}= E(X_{\rm [110]}) -
E(X_{\rm [1\bar{1}0]})$.

\begin{figure}
\begin{center}
\includegraphics[width=3.0in,angle=0]{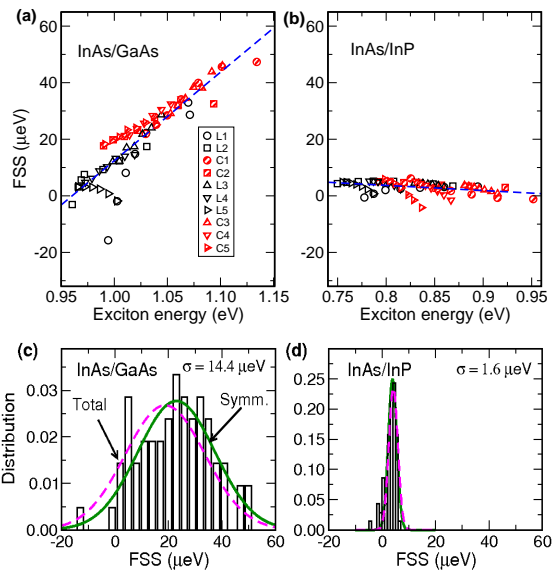}
\end{center}
\caption{The upper panel shows 
the FSS vs the exciton energy for (a) InAs/GaAs and (b) InP QDs.
The lower panel shows the FSS distributions for (c) InAs/GaAs and (d) InP QDs.
The solid lines are fitted by Gaussian functions for all shape-symmetric
QDs, whereas the
dashed lines represent the distributions of the FSS of total 
samples including also the asymmetric dots. $\sigma$ is the standard deviation
of the distribution.}
\label{fig:fssvsenergy} 
\end{figure}

\begin{figure}
\begin{center}
\includegraphics[width=3.2in,angle=-90]{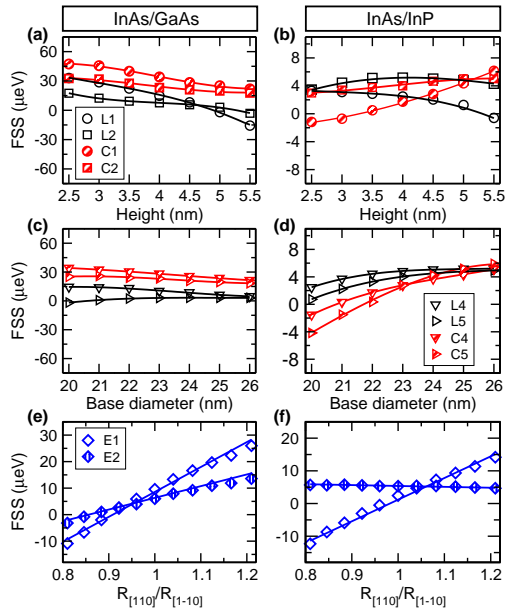}
\end{center}
\caption{
Upper panel:The height dependence of the FSS for (a) InAs/GaAs and
(b) InAs/InP QDs with different sizes and geometries (Lens and Cone). 
Middle panel: The base diameter dependence of the FSS for (c) InAs/GaAs and (d)
InAs/InP QDs. 
Lower panel:The FSS as function
of lateral aspect ratio of the elongated QDs in (e) InAs/GaAs and (f)
InAs/InP QDs.
}
\label{fig:geomeffect}
\end{figure}

Figure \ref{fig:fssvsenergy} shows the FSS in these 
two systems as a function of the excitonic energy\cite{compareFSS}.
The exciton energies of the (pure) InAs/GaAs QDs range from 0.95 - 1.12 eV, 
and the FSS scatters from - 18 $\mu$eV to 60 $\mu$eV, 
both are in a good agreement with experiments
\cite{young05,hogele04,akopian06, tartakovskii04,seguin05},
which establishes credibility for the
results of the InAs/InP dots.
Figure \ref{fig:fssvsenergy} demonstrates that the intrinsic FSS of the
InAs/InP QDs (-4 $\mu$eV to 6 $\mu$eV), is about
an order of magnitude smaller than that of the InAs/GaAs QDs. 
This system can
therefore be used as efficient entangled photon source. Furthermore, since the
primary exciton wavelength of the InAs/InP QDs is around 1.55 $\mu$m telecom
wavelength\cite{cade06,kim05, gong08a}, they are very promising for the long
distance quantum communication via optical fibers.
The calculated FSS are consistent with recent
experiments, in which extremely small exciton FSS were
measured in the InAs/InP dots\cite{cade06,kim05}. 
However, it was not realized that the small FSS
 is an {\it intrinsic} property of the InAs/InP dots.

The FSS of the InAs/GaAs and InAs/InP QDs are compared in 
Fig. \ref{fig:geomeffect} for different
dot geometries given in Table I . Figure I(a)-(d) show the results for
shape-symmetric dots (circular-base), whereas parts (e)-(f) illustrate the
effects of shape asymmetry (base with different radii lengths). We next
discuss the trends with dot height, base size, and shape anisotropy:

{\it Intrinsic FSS vs dot height:}
For the InAs/GaAs dots, we see the FSS decrease monotonically with increasing 
of the dot height. At about $h$ = 5 nm, the FSS of the InAs/GaAs dot L1 
is found to be zero. This is because in the tall InAs/GaAs QDs,
the holes are localized
at the interface of the dot due to the strain effect \cite{he04a},
which reduces the electron-hole wavefunction overlap, 
leading to small FSS. 
However, it is not a good way to reduce the FSS by increasing the dot 
height for the InAs/GaAs QDs, because the PL intensity is also reduced 
with the deduction of the FSS
due to the reducing of electron-hole overlap.
The height effect of the FSS is less dramatic for the InAs/InP dots, 
and all FSS are 
found to be extremely small(between -4 $\mu$eV to 6 $\mu$eV).
A zero FSS is also found for the L1 InAs/InP dots at height $\sim$ 5.5 nm.
However, unlike in the InAs/GaAs dot, the holes are not localized on
the interface in the InAs/InP QDs \cite{gong08a},
therefore, it would not suffer the problem of PL intensity suppression
as in the tall InAs/GaAs dots.

{\it Intrinsic FSS as a function of dot base size:} 
For flat InAs/GaAs QDs (L4, C4), the FSS decrease monotonically with
increasing of the base size. However, for the tall InAs/GaAs dot(L5), as we
increase the base size, the FSS increases, because the hole is less localized
on the interface \cite{he04a}, which increases the electron-hole overlap, and thus
the FSS. However, the FSS depends less on the base diameters than on the dot
height for the InAs/GaAs dots. For InAs/InP dots, the FSS increase slightly as
we increase the base size for all dot geometries. The FSS of
the cone-shaped dots are similar to that of the lens-shaped dots.

{\it Effect of Shape-asymmetry to FSS :} 
Figure \ref{fig:geomeffect}  (e) 
depicts the FSS of the InAs/GaAs QDs as functions of the 
lateral aspect ratio  $R_{[110]}/R_{[1\bar{1}0]}$, 
whereas Fig. \ref{fig:geomeffect} (f) shows the results for 
the InAs/InP QDs, for two fixed dot volume and height. 
For InAs/GaAs QDs, we see that as we increase the asymmetric ratio 
the FSS increase dramatically from -11 $\mu$eV to 26 $\mu$eV. 
For the InAs/InP QDs, we found that for the smaller QDs, 
the FSS depends strongly on the shape asymmetry 
of QDs (from -12 $\mu$eV to 14 $\mu$eV ),
whereas for the larger dot, 
the FSS only weakly dependent on the shape-asymmetry.

{\it Statistical distribution of FSS vs sizes and shapes :}
To further illustrate the differences 
of the FSS between the InAs/GaAs and the InAs/InP QDs,
we plot the statistical distribution of FSS for the two types of dots in 
Fig. \ref{fig:fssvsenergy}(c),(d) respectively.  
The solid lines represent the distributions of the intrinsic FSS of 
cylindrical QDs, fitted by Gaussian functions
and the dashed line represent the distributions of the FSS of 
total samples including also the asymmetric dots. As we see, including the
asymmetric dots does not change much the FSS distribution.
The mean value of the FSS of the InAs/GaAs dots is 23 $\mu$eV, with standard
deviation of 14.4 meV, whereas the average 
FSS of the InAs/InP dots is 3.5 $\mu$eV and standard deviation 
is only about 1.6 $\mu$eV.

The FSS of the InAs/InP might still be too 
large to achieve perfect photon entanglement \cite{stevenson06}, 
but it can be further reduced by controlling the growth condition. For
example, for the ``small'' InAs, InP lattice mismatch, 
one may try to grow a quantum disk, which has higher D$_{2d}$ symmetry, 
thus (almost) zero FSS. Nevertheless,
it is much easier to obtain the InAs/InP QDs with nearly zero FSS than the
InAs/GaAs dots.

To conclude, we have shown that the {\it intrinsic} 
FSS of the InAs/InP dots
are (statistically) about an order of magnitude 
smaller than that of the InAs/GaAs dots.  
The InAs/InP QDs have additional advantages because the 
emission photon wavelength is around 1.55 $\mu$m, and is far away from the
wetting layer background emissions. Combining these advantages, 
one can expect that the InAs/InP QDs can play a crucial role in the quantum
information applications, as a new generation of
``on-demanding'' entangled photon source.



L.H. acknowledges the support from the Chinese National
Fundamental Research Program 2006CB921900, the Innovation
funds and ``Hundreds of Talents'' program from Chinese Academy of
Sciences, and National Natural Science Foundation of China (Grant
No. 10674124). A.Z. acknowledges support from USA DOE Office of Science,
Basic Energy Science, Materials Sciences and Engineering, LAB-17 initiative,
under Contract No. DE-AC36-99GO10337 to NREL.


\begin{thebibliography}{25}
\bibitem[{com()}]{compareFSS}
\bibinfo{note}{In present work, we show the {\it intrinsic} FSS for selected
  geometries, which should not be directly compared to the trend of the FSS vs
  exciton energy for a particular thermal annealing process, (e.g Ref.
  \onlinecite{young05}), because the structure and composition of the QDs
  change dramatically during the thermal annealing.}
\end{thebibliography}

\end{document}